\documentclass[fleqn,usenatbib]{mnras}

\usepackage[T1]{fontenc}

\DeclareRobustCommand{\VAN}[3]{#2}
\let\VANthebibliography\thebibliography
\def\thebibliography{\DeclareRobustCommand{\VAN}[3]{##3}\VANthebibliography}

\usepackage{graphicx}	
\usepackage{amsmath}	
\usepackage{amssymb}	
\usepackage{subfig}
\usepackage[dvipsnames]{xcolor}
\usepackage{soul}
\usepackage{newtxtext,newtxmath}
\usepackage[symbol]{footmisc}

\newcommand{\kms}{\,km\,s$^{-1}$}	
\newcommand{\msun}{${\rm M}_{\odot}$\,}

\newcommand{\Ion}[2]{#1\,{\sc #2}}

\setstcolor{red}

\title[The donor radial velocity in GY Cnc]{The donor star radial velocity curve in the cataclysmic variable GY~Cnc confirms white dwarf eclipse modelling mass}

\author[S. P. Littlefair et al]{
S. P. Littlefair$^{1}$\thanks{E-mail: s.littlefair@shef.ac.uk},
Pablo Rodr\'{i}guez-Gil$^{2, 3}$,
T. R. Marsh$^{4, \dag}$,
S. G. Parsons$^{1}$
and V. S. Dhillon$^{1, 3}$
\\
$^{1}$Dept of Physics and Astronomy, University of Sheffield, Sheffield, S3 7RH, UK\\
$^{2}$Instituto de Astrof\'{i}sica de Canarias, E-38205 La Laguna, Tenerife, Spain\\
$^{3}$Departamento de Astrof\'{i}sica, Universidad de La Laguna, E-38206 La Laguna, Tenerife, Spain\\
$^{4}$Department of Physics, University of Warwick, Coventry CV4 7AL, UK\\
}

\date{Accepted XXX. Received YYY; in original form ZZZ}

\pubyear{2022}

\begin{document}
\label{firstpage}
\pagerange{\pageref{firstpage}--\pageref{lastpage}}
\maketitle

\begin{abstract}
A large number of white dwarf and donor masses in cataclysmic variables have been found via modelling the primary eclipse, a method that relies on untested assumptions. Recent measurements of the mass of the white dwarf in the cataclysmic variable GY~Cnc, obtained via modelling its ultraviolet spectrum, conflict with the mass obtained via modelling the eclipse light curve. Here we measure the radial velocity of the absorption lines from the donor star in GY~Cnc to be $K_{\rm abs} = 280 \pm 2$\,\kms, in excellent agreement with the prediction based on the masses derived from modelling the eclipse light curve.
It is possible that the white dwarf mass derived from the ultraviolet spectrum of GY~Cnc is affected by the difficulty of disentangling the white dwarf spectrum from the accretion disc spectrum.\end{abstract}




\section{Introduction}
Cataclysmic variables (CVs) are interacting binaries in which a white dwarf accretes material from a low-mass companion (the donor star). CVs are a valuable test-bed for binary evolution, since their long-term evolution is driven by angular momentum loss from the binary. For an in-depth review of the evolution of CVs, see \cite{knigge11b}. 

The current predictions of CV evolution \cite[e.g.][]{kolb93, howell01} are in tension with the high average white dwarf masses in CVs \citep{zorotovic2011} and the scarcity of CVs that have evolved past the period minimum \citep{pala2020}. There are also issues with the generally accepted explanation for the period gap \citep{andronov2003, garraffo2018, elbadry2023}. Solutions to many of these problems may be found by invoking additional sources of angular momentum loss from the binary. For example, \cite{2016MNRAS.455L..16S} suggest that additional angular momentum loss caused by binary interactions with novae ejecta could solve many of the outstanding issues with CV evolution, and theoretical models of novae outflows seem to support such a mechanism \citep{2022ApJ...938...31S}.

The cause and magnitude of angular momentum loss from CVs is therefore quite uncertain, and likely to be key to understanding their evolution. This makes observational constraints important. The best observational constraints on the angular momentum loss rates come from measurements of the {\em long term} mass transfer rate, $\dot{M}$, provided these measures average over longer timescales than any possible short term fluctuations \citep{knigge11b}. Two methods satisfy this requirement:
White dwarf effective temperatures measure the mass transfer rate averaged over $\sim10^4$\,yr \citep{Townsley_Bildsten_2003, knigge11b, pala2022}. Donor star radii measure an average of the mass transfer rate over $\sim 10^8$\,yr \citep{knigge11b}, making them potentially the best test of secular mass transfer rates.

Precise measures of the masses and radii of the donors and white dwarfs in CVs are difficult to obtain. This is because the accretion disc often outshines the donor star and white dwarf, so CVs very rarely present as double-lined eclipsing binaries. In some cases, the white dwarf is visible in the ultraviolet, and spectroscopic modelling can yield the white dwarf mass and temperature \cite[e.g.][]{godon2022, pala2022}. However, the majority of precise component masses for CVs arise from modelling of the shape of the primary eclipse, hereafter called the {\em photometric eclipse method} \cite[e.g.][]{feline05, littlefair06, savoury11, mcallister2019}.

Masses derived from the photometric eclipse method have small statistical errors, and have been independently confirmed in a few systems where it is also possible to measure radial velocities via spectroscopy \citep{tulloch09, copperwheat2012, savoury12}. However, they are underpinned by untested assumptions. In particular, to infer the white dwarf radius from the ingress/egress duration, it is assumed that the whole white dwarf is visible and that the white dwarf surface brightness is uniform (apart from limb darkening). There are good reasons to suspect this is not true in all CVs. Firstly, the presence of gas between the observer and white dwarf surface is well established. This gas is revealed by the so-called ``iron curtain" -- a forest of absorption lines in the blue and ultraviolet spectra of CVs \citep{horne94}. It is quite possible that absorption from the iron curtain could mean that we do not see the whole white dwarf surface. In addition, a boundary layer exists near the white dwarf surface, where the rapidly rotating disc material transforms kinetic energy into heat and radiation. It is possible that the "white dwarf ingress" seen during eclipse is in fact dominated by the boundary layer itself and not the white dwarf surface. \cite{spark2015} present high-cadence light curves of white dwarf eclipses that appear to show departures from the expected shape of a limb-darkened white dwarf. These departures may be the result of random flickering, but they may also represent a breakdown in the assumptions underpinning the photometric eclipse method.

\cite{godon2022} present a recent independent test of photometrically derived white dwarf masses. Using \textit{Gaia} eDR3 parallaxes, and archival {\em Hubble Space Telescope} ultraviolet spectroscopy, they model the white dwarf and the iron curtain simultaneously and derive a white dwarf mass directly from the ultraviolet spectrum. Five of the systems modelled have independent masses derived from the eclipse light curve, and for four of the five systems excellent agreement is found. However, the fifth system, GY~Cnc \citep{gaensicke2000}, is discrepant. \citeauthor{godon2022} found a white dwarf mass of $M_1 = 0.57 \pm 0.05$~\msun, compared to $M_1 = 0.88\pm0.02$~\msun obtained from modelling the eclipse light curve \citep{mcallister2019}, and they also found that modelling GY~Cnc required a significant iron curtain, with a hydrogen column density of $5\times10^{21}$\,cm$^{-2}$. This raises the possibility that the  masses in GY~Cnc derived from eclipse modelling are incorrect and therefore brings into question all masses derived via eclipse modelling that do not have independent confirmation.

 In this paper, we present archival spectroscopy of GY~Cnc that reveals the orbital motion of the donor star. We measure the velocity of the donor star's absorption lines, and determine if it is consistent with the system parameters measured by \cite{mcallister2019} from the eclipse light curve.

\section{Observations and Data Reduction}

GY Cnc was observed on the night of 14 April 2006 using ISIS on the 4.2-m William Herschel Telescope (WHT) at La Palma, Spain, as part of programme NL/2006A/5 (PI: Ham). Note that in this study, we only analyse data from the red arm of ISIS: the R316R grating used with a 1-arcsec slit and a central wavelength of 6304 \AA. The detector was binned by a factor of 2 in the spatial and spectral directions, giving an unvignetted wavelength coverage of 5400 -- 7950\,\AA\, and a resolution of approximately 3.8 \AA. The slit was rotated to a sky position angle of 101$^{\circ}$ to include the spectrum of Gaia DR3 635520611367973120 on the slit, to help measure the telluric contamination of the GY~Cnc spectra. A total of 219 exposures with an exposure time of 50\,s were taken, but the first 20 and the last 14 spectra were affected by clouds and were discarded. The resulting time series covers from UT~20:30 to 00:30. Frequent CuNe + CuAr arcs were taken during the night to ensure accurate wavelength calibration.

The raw data were reduced, and spectra of both objects extracted using the \textsc{python} package {\sc aspired} \citep{2020arXiv201203505L}. Raw images were de-biased and corrected for uneven slit illumination and pixel response non-uniformity using observations of a tungsten lamp. The spatial positions of the spectra were traced with a third order polynomial and spectra were extracted using an aperture of width 12 pixels and the optimal extraction method of \cite{horne86a}. The sky was subtracted using a first order polynomial fit to pixel values in a pair of strips of width 6 pixels, separated by 5 pixels from the edges of the object aperture.

Wavelength calibration for each spectrum was carried out using the arc spectrum taken nearest in time. A second order polynomial was used to fit the relation between pixel number and wavelength with root-mean-square residuals of order 0.4 \AA, which is less than a single pixel. The spectra were flux calibrated using observations of the DA7 white dwarf Gaia DR3 4452521234885949184.

\begin{figure}
    \centering
    \includegraphics[width=0.5\textwidth]{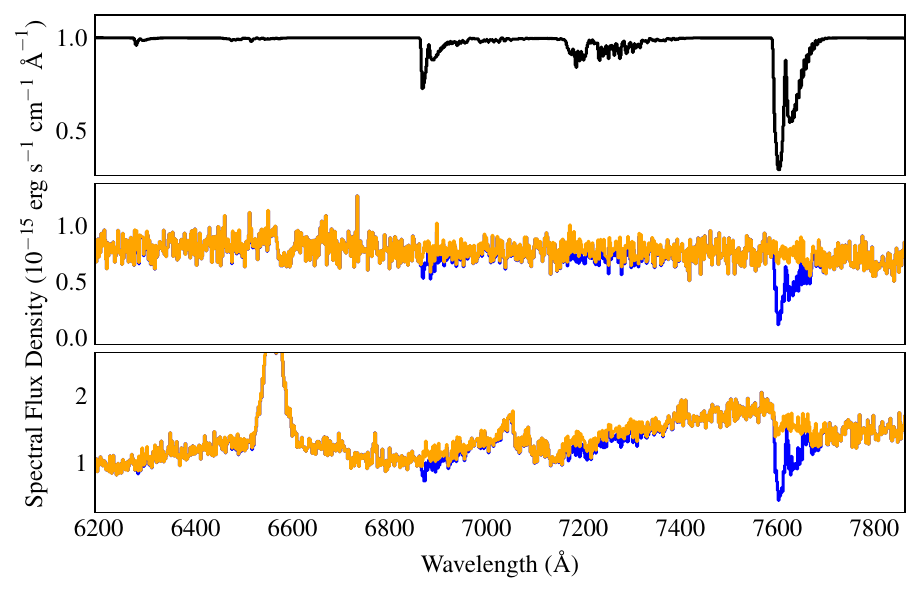}
    \caption{Telluric correction of spectra. Top: The best-fit theoretical telluric spectrum found by fitting the LBLRTM to the spectrum of Gaia DR3 635520611367973120. Middle: an individual spectrum of Gaia DR3 635520611367973120 (blue) and the same spectrum after correcting with the theoretical telluric spectrum above (orange). Bottom: the spectrum of GY Cnc, zoomed in to focus on  the donor star features before (blue) and after (orange) telluric correction.}
    \label{fig:telfit}
\end{figure}

Telluric correction was carried out using the \textsc{python} package {\sc telfit} \citep{2014AJ....148...53G}. {\sc telfit} calculates a theoretical telluric spectrum using the Line By Line Radiative Transfer Model \citep[][LBLRTM]{clough2005}, and optimises the atmospheric parameters to find the telluric spectrum that best matches the observed spectrum. We calculated the telluric spectrum using the relatively featureless spectrum of Gaia DR3 635520611367973120 (K4V), and used this theoretical telluric spectrum to correct the spectra of GY Cnc. An example is given in Fig.~\ref{fig:telfit}.

\section{Results}
\begin{figure}
    \centering
    \includegraphics[width=0.5\textwidth]{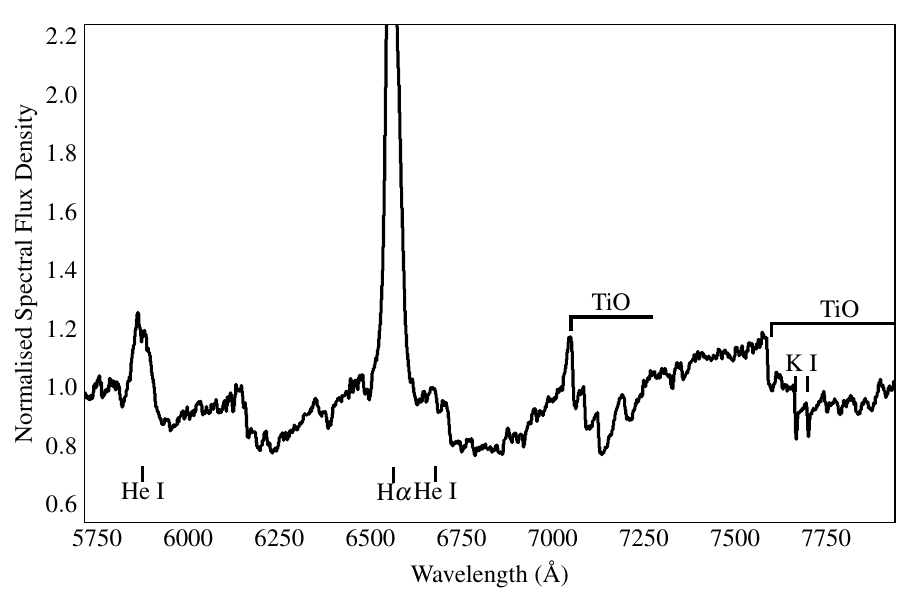}
    \caption{Average spectrum of GY Cnc cropped to focus on the donor star features. The spectrum is presented in the rest frame of the donor star, using the radial velocity found in section~\ref{subsec:rv}. Absorption bands of TiO and the \Ion{K}{i} doublet from the donor star are visible along with strong, double-peaked emission lines from the disc. The continuum shape has been removed with a second order polynomial fit.}
    \label{fig:av}
\end{figure}

Analysis of the reduced spectra was carried out using {\sc molly}\footnote{{\sc molly} was written by T.\ R.\ Marsh and is available at \url{http://www. warwick.ac.uk/go/trmarsh/software}}. All spectra were corrected for the velocity shift of the observer with respect to the Solar System Barycenter and binned onto an identical wavelength scale with bins uniformly spaced in velocity units. 
Spectra were normalised by dividing by a second order polynomial fit, and cosmic rays were iteratively removed from the spectra by clipping pixels whose values lay more than 5$\sigma$ away from the mean value of the equivalent pixels in the entire time series. The average spectrum of GY Cnc in the rest frame of the donor star (using the radial velocity found in section~\ref{subsec:rv}) is shown in Fig.~\ref{fig:av}.

Because the spectrum of GY~Cnc shows strong absorption features from the donor star, we can use them to measure its radial velocity and test which of the two claimed white dwarf masses for GY~Cnc is correct.

\subsection{Trailed spectra}
\begin{figure*}
  \includegraphics[width=\textwidth]{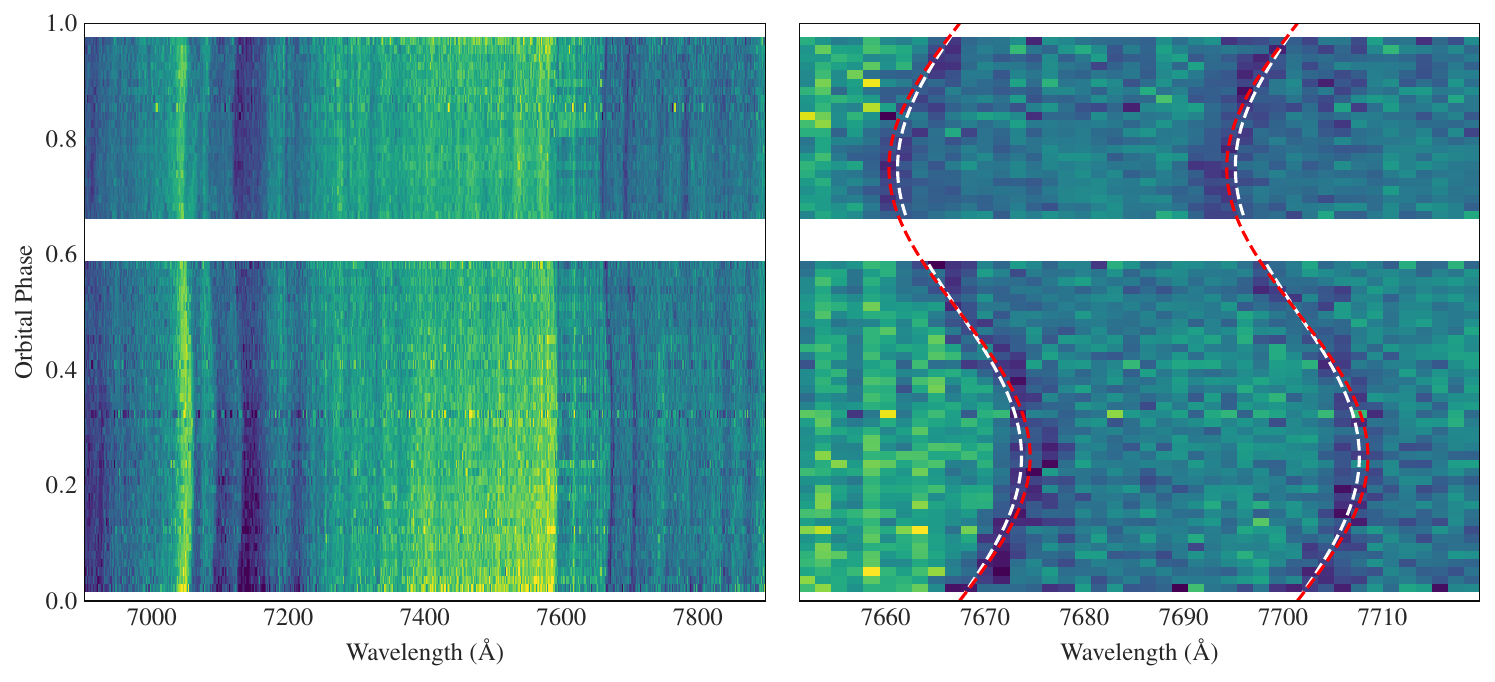}
  \caption{The normalised trailed spectra of GY Cnc, focusing on the donor star features redward of 6900\,\AA. The left panel shows the strong TiO bandheads at 7050 and 7600\,\AA, which clearly trace the donor star's radial velocity. Vertical stripes, and the noisy region around 7600\,\AA\ represent imperfect subtraction of telluric features. The right-hand panel shows a zoomed region around the \ion{K}{i} doublet at 7665/7699\,\AA. Superimposed are dashed lines showing two potential solutions for the radial velocity of the donor star, one with 244\,\kms\ (white) and one with 278\,\kms\ amplitude (red). These two velocities represent the estimated donor star velocity assuming the white dwarf masses of 
  \protect\cite{godon2022} and \protect\cite{mcallister2019}, respectively.}
  \label{fig:trail}
\end{figure*}

Figure~\ref{fig:trail} shows the trailed spectra of GY~Cnc, where each spectrum is represented as a row in the image, and the brightness of the image represents the spectral flux density. 
Orbital phases for each spectrum were calculated using the ephemeris of \cite{mcallister2019}. The trailed spectra reveal that the TiO bandheads and the \Ion{K}{i} absorption lines trace the motion of the donor star, but also show that our telluric correction is not perfect. Residual telluric lines form vertical streaks in the trailed spectra and the residuals from the strong telluric region contaminate the TiO bandhead at 7600\,\AA. However, the TiO bandhead at 7050\,\AA\ and the \Ion{K}{i} doublet at 7680\,\AA\ remain largely free of contamination from telluric features, and should therefore be suitable for determining the radial velocity of the donor star. Indeed, visual inspection of the motion of the \Ion{K}{i} lines in Fig.~\ref{fig:trail} appears to support a radial velocity amplitude for the donor star of $\simeq 280$\,\kms, larger than the prediction assuming the white dwarf mass of \protect\cite{godon2022} and comparable to the value predicted using the white dwarf mass of \protect\cite{mcallister2019} (see section~\ref{sec:discussion} for more details).

\subsection{Radial velocity curve of the donor star}
\label{subsec:rv}
We derive the amplitude of the radial velocity curve of the donor star's absorption lines, $K_{\rm abs}$, by cross correlating the normalised spectra of GY~Cnc against a solar-metallicity M3\,V spectral template from \cite{kesseli2017}. These templates are produced from Sloan Digital Sky Survey (SDSS) spectroscopy with a resolution of $R \approx 2000$, which is well matched to our data. Templates with spectral types from M1\,V to M5\,V (see Section~\ref{subsec:donor_spectype}) were trialled, with no significant difference in the results.

\begin{figure}
    \centering
    \includegraphics[width=0.5\textwidth]{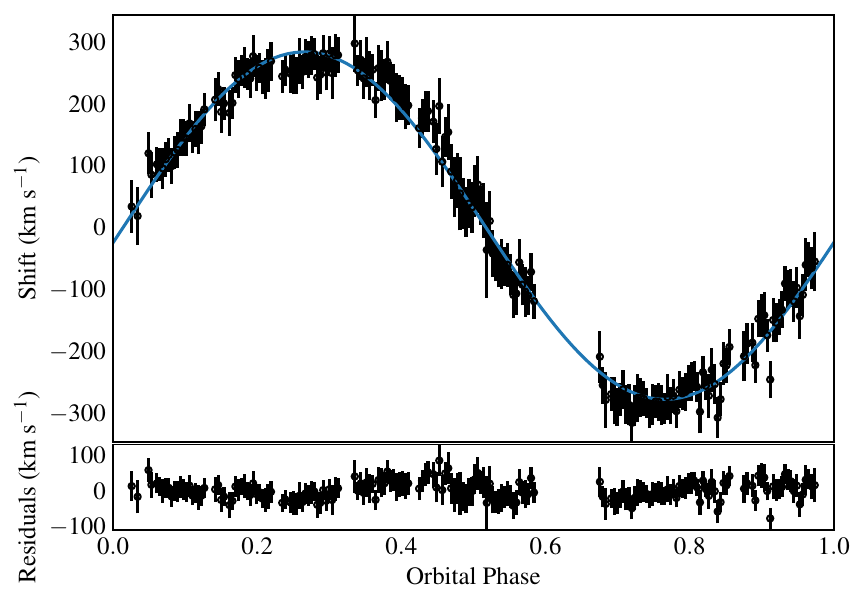}
    \caption{Radial velocities for the absorption features in GY Cnc, measured via cross-correlation and plotted as a function of orbital phase. Points with uncertainties show the radial velocity measurements. The blue line shows the best-fit sinusoid.}
    \label{fig:rvs}
\end{figure}

The spectra were first cropped to exclude data outside the range 6750--7900 \AA, and normalised by dividing by their means. Following this, a first order polynomial fit was subtracted from both spectrum and template, and both spectra were resampled onto the same wavelength scale, which is uniformly spaced in log-wavelength. Prior to cross-correlation, both spectra were apodized by multiplying by a cosine function that is restricted to the first and last 5 per cent of the spectrum; this reduces the impact of edge effects. Cross-correlation was performed on the flux data of the two spectra, weighted by the uncertainties on the spectrum of GY~Cnc. From the cross-correlation of GY~Cnc and template, radial velocities and their uncertainties were found by fitting a parabola to the three points nearest the peak of the cross-correlation. Radial  velocities are shown in Fig.~\ref{fig:rvs}. 

Orbital phases for each spectrum were calculated using the ephemeris of \cite{mcallister2019}, and a sinusoid was fitted to the radial velocities:
\begin{equation}
    V(\phi) = V_{\rm sys} + K_{\rm abs} \sin[2\pi (\phi + \phi_0)]~,
\end{equation}
\noindent
with $\phi$ the orbital phase and $V_{\rm sys}$ the systemic velocity. $\phi_0$ is an offset in phase that allows for the possibility that the radial velocity of the donor is not zero at mid-eclipse. The best-fit solution has $K_{\rm abs} = 280 \pm 2$\,\kms\ , $\phi_0 = -0.02 \pm 0.02$ and $V_{\rm sys} = 3 \pm 2$\,\kms\ and is shown in Fig.~\ref{fig:rvs}. The residuals to our fit indicate systematic deviations from a pure sinusoid that may reveal the effects of irradiation on the absorption line strength, or the effects of starspots. However, the solution has a reduced chi-squared of $\chi^2_{\nu} = 0.54$, indicating that the uncertainties on the radial velocities are slightly overestimated. These deviations are therefore unlikely to significantly affect our estimate of $K_{\rm abs}$.

\subsection{Donor star spectral type}
\label{subsec:donor_spectype}
\begin{figure}
    \centering
    \includegraphics[width=0.5\textwidth]{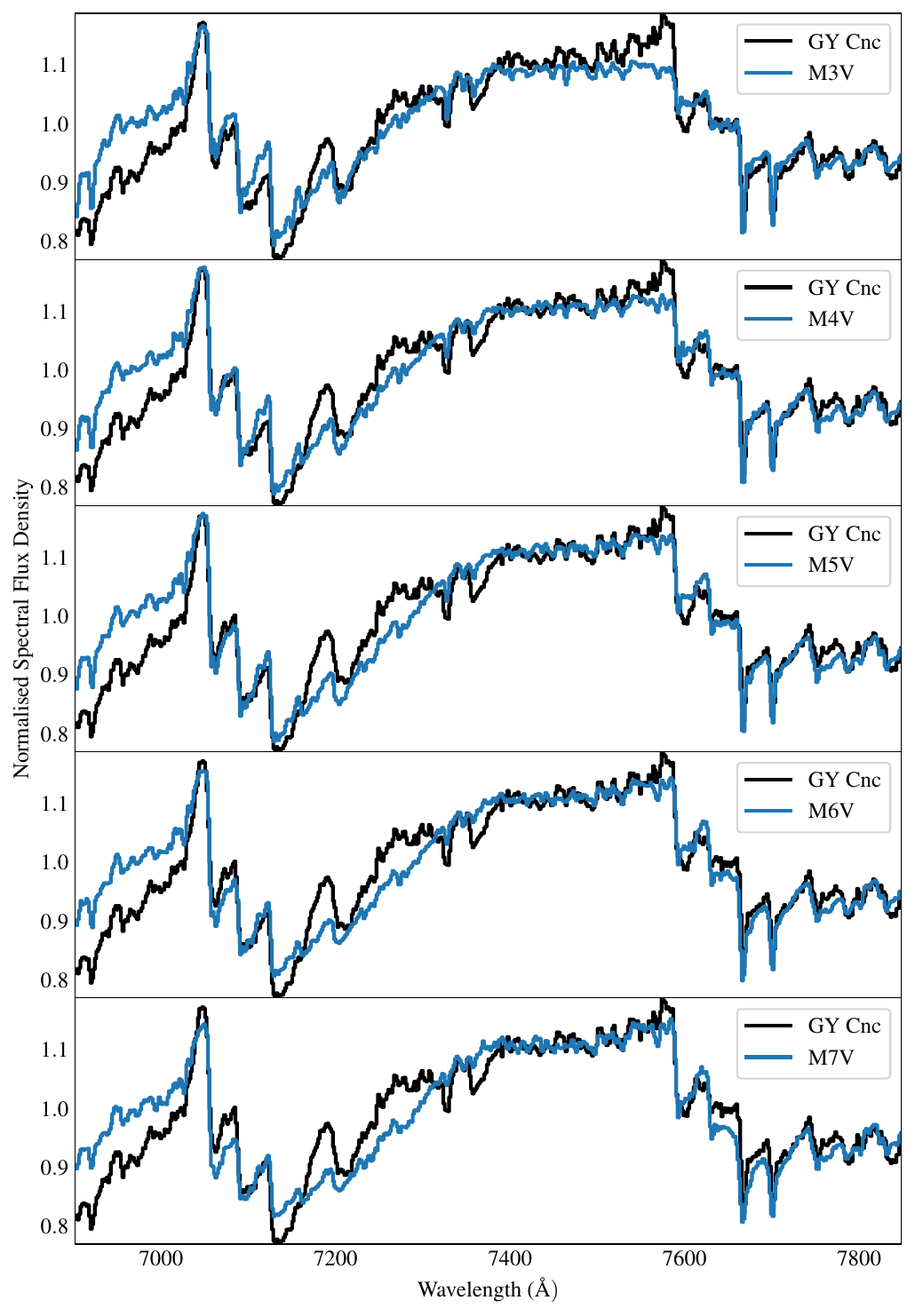}
    \caption{Comparison of the spectrum of GY~Cnc to M-dwarf template spectra. The spectrum of GY~Cnc in the rest frame of the donor star is shown in black, and template spectra are shown in blue. Spectra were normalised by dividing by the overall mean flux. See text for details. }
    \label{fig:optsub}
\end{figure}

Using $K_{\rm abs} = 280$\,\kms, we created an average spectrum of GY~Cnc in the rest frame of the donor star, and compared to the solar-metallicity library templates of \cite{kesseli2017}. The templates and GY~Cnc spectra were normalised using a first order polynomial fit and resampled onto the same wavelength scale. For each template, we added a range of constant values to simulate contamination by accretion disc light, and re-normalised. The best-fit template and contamination levels were found by subtracting the template from the GY~Cnc spectrum to create a residual spectrum. We then calculate the $\chi^2$ of the residual spectrum, and a smoothed version of itself (Gaussian smoothing, $\sigma=10$\,pixels). This reduces the impact of continuum shape on the figure of merit, which is dominated by the strengths of the TiO and \Ion{K}{i} absorption features. A comparison between GY~Cnc and the best-fit spectrum at each spectral type is shown in Fig.~\ref{fig:optsub}. The best-fit template has a spectral type of M5\,V. This is slightly later than suggested by the semi-empirical donor star sequence of \cite{knigge11b}, which predicts a spectral type of M3\,V at the orbital period of GY~Cnc. Based on visual inspection, \cite{thorstensen2000} find a spectral type of M3\,V. The M3\,V template is a much poorer fit to our data ($\chi^2 = 2932$, 568 D.O.F) than the M5\,V template ($\chi^2 = 2578$, 568 D.O.F). This is largely driven by the strength of the TiO bandhead at 760\,nm.

\subsection{Radial velocity of the emission line wings}
\label{subsec:rvha}
\begin{figure}
    \centering
    \includegraphics[width=0.5\textwidth]{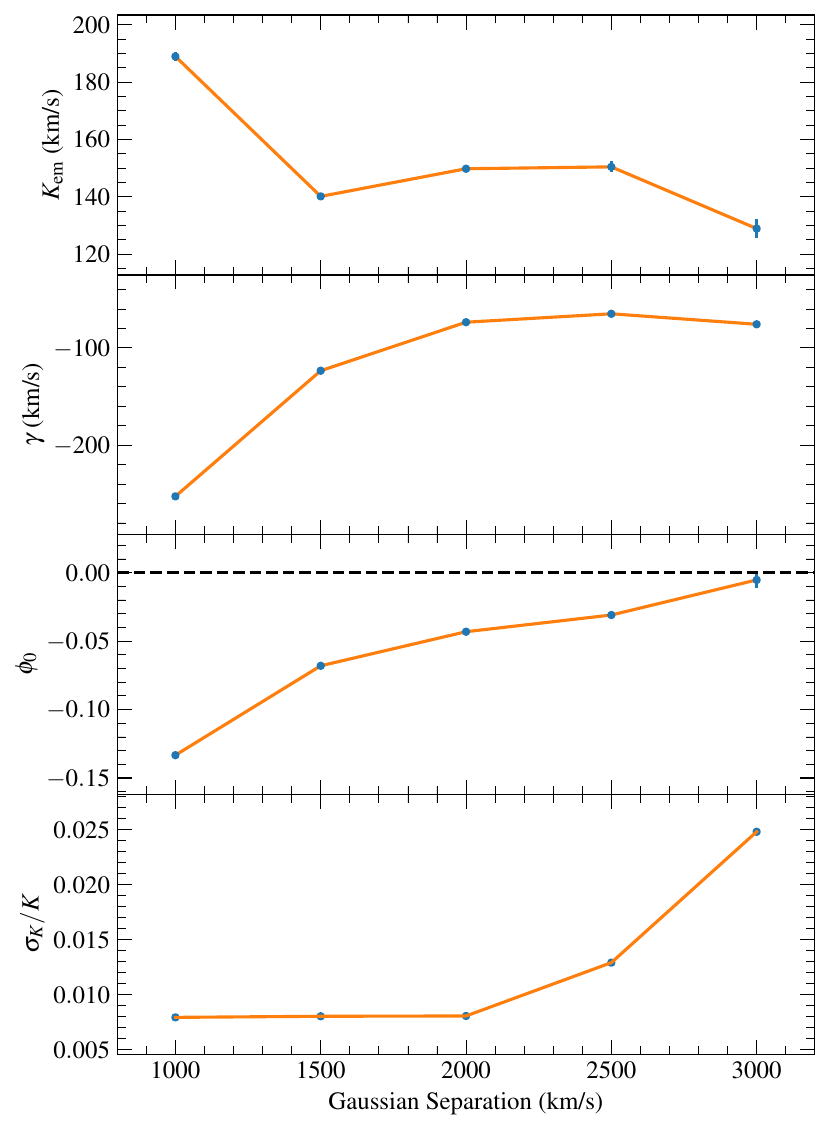}
    \caption{Diagnostic diagram for the H$\alpha$ emission line.}
    \label{fig:diagnostic}
\end{figure}
It is not possible to directly measure the radial velocity of the white dwarf, as no spectral features are visible. Historically, mass determinations in CVs have attempted to estimate the white dwarf radial velocity via the motion of the emission line wings. The motion of the line wings can be determined using the method of \cite{schneider80}. An emission line profile is convolved with a double-Gaussian profile, each Gaussian having the same width but opposite amplitudes. When the Gaussians are centred on the emission line, the convolution vanishes, determining the radial velocity. At large separations of the Gaussians, one measures the velocity of the emission line wings, formed in the high-velocity gas near the white dwarf. It is hoped that the motion of this gas traces the white dwarf itself, although whether this is true is far from certain.

We measure the velocities of the H$\alpha$ line wings using this approach. For a given separation of the Gaussians, we fit a sinusoidal radial velocity curve:
\begin{equation}
    V(\phi) = V_{\rm sys} - K_{\rm em} \sin[2\pi (\phi + \phi_0)]~,
\end{equation}
\noindent
A diagnostic diagram, showing the fit parameters of this curve versus the Gaussian separation is shown in figure~\ref{fig:diagnostic}. With a Gaussian separation of 3000\,\kms we find $K_{\rm em} = 129 \pm 3$\,\kms, $\phi_0 = -0.05 \pm 0.06$ and $V_{\rm sys} = -76 \pm 3$\, \kms. These uncertainties do not include the systematic uncertainty arising from choosing an appropriate Gaussian separation, which can change the measured velocities by tens of \kms.

Whilst we strongly caution against interpreting $K_{\rm em}$ as the velocity of the white dwarf, we note that our value is not too far away from the value measured by \cite{thorstensen2000} of $K_{\rm em} = 116 \pm 5$\, \kms from the same line, and is in agreement with the predicted white dwarf velocity $K_{\rm wd} = 125 \pm 4$\, \kms of \cite{mcallister2019}.

\section{Discussion}
\label{sec:discussion}
Based on modelling of the eclipse light curve, and a theoretical white dwarf mass-radius relation, \cite{mcallister2019} predict a donor star radial velocity amplitude of $K_2 = 278\pm 2$\,\kms\ for GY\,Cnc. This compares very well with the velocity we measure for the absorption lines, $K_{\rm abs} = 280 \pm 2$\,\kms, and indicates a preference for the system parameters of \cite{mcallister2019} over those of \cite{godon2022}. However, care must be taken, since the centre of light of the donor absorption lines does not necessarily correspond to its centre of mass \citep[e.g.][]{watson2003}. In particular, irradiation from the white dwarf can strengthen or weaken absorption lines near the L1 point, so that $K_{\rm abs} \ne K_2$. Non-uniform light distribution on the donor star can be detected in the radial velocity curves as a departure from a sinusoidal fit. Our data lack the precision to see this effect and thus we cannot quantify any difference between $K_{\rm abs}$ and $K_2$. 

In the case of GY~Cnc, the spectral type of M5\,V means that the TiO bandheads will weaken with increasing temperature. Thus, these features are expected to be weaker near the L1 point if this region is significantly hotter (see Fig.~\ref{fig:optsub}). Therefore, it is reasonable to assume that in GY~Cnc the centre of light will be shifted towards the back half of the donor star, implying $K_{\rm abs} \ge K_2$. Since higher values of $K_2$ correspond to larger white dwarf masses, we are confident that the white dwarf mass of \cite{mcallister2019} should be preferred. We therefore asked ourselves if it is possible to use additional constraints to {\em rule out} the white dwarf mass of \cite{godon2022}.

 A detailed study of the donor absorption lines and radial velocity of the emission line wings in GY~Cnc was carried out by \cite{thorstensen2000}. Their measured $K_{\rm abs} = 297 \pm 15$\,\kms agrees with our value, within the uncertainties. To derive the white dwarf mass from their data the author assumed the wings of the emission lines trace the velocity of the white dwarf, and applied ``a frank guess''  for the correction from $K_{\rm abs}$ to $K_2$. The result is a white dwarf mass of $M_1 = 0.82 \pm 0.14$~\msun, which favours the higher white dwarf mass of \cite{mcallister2019}, but is only marginally inconsistent (1.7$\sigma$) with the lower mass of \cite{godon2003}. Without the correction to $K_{\rm abs}$, \cite{thorstensen2000} find $M_1 = 0.99 \pm 0.12$~\msun, which is inconsistent with the mass of \cite{godon2003}. 

The component masses of a CV can be derived without assuming the emission lines reveal the white dwarf's velocity using measurements of the donor star's absorption lines and the phase width of the white dwarf eclipse, $\Delta \phi_{1/2}$ \citep[e.g.][]{smith98}. However, this requires measurement of the rotational broadening of the donor star's absorption lines, $V_{\rm rot}\sin i$, and our spectra lack the resolution for this. For the system parameters in \cite{mcallister2019}, we would expect $V_{\rm rot}\sin i \approx 125$\,\kms, which corresponds to a broadening of less than one resolution element. Since we cannot measure $V_{\rm rot}\sin i$, we can either adopt $q=0.448$ from the light curve fitting \cite{mcallister2019}, or we can adopt $q=K_{\rm em}/K_{\rm abs} = 0.46$ from the spectroscopic data (we do not include errors due to the difficulty assigning systematic errors to $K_{\rm em}$). Because the value of $K_{\rm em}$ can change markedly depending on the Gaussian separation used, and because it is far from certain that the gas in the inner disc traces the radial velocity of the white dwarf, we prefer not to use $K_{\rm em}$ to measure the system parameters if an alternative is available. 

We will therefore proceed by adopting the most robust parameters from the light curve fitting of \cite{mcallister2019}, although we note that our conclusions do not change, whichever value we adopt. Criticisms of the eclipse light curve modelling method mostly focus on inferring the white dwarf radius, and hence mass, from the ingress and egress widths of the white dwarf eclipse. Constraints upon the mass ratio, $q$, are independent of this effect, as they come largely from the timing of the bright spot eclipse. We assume here that $\Delta \phi_{1/2}$ is also robust. In principle, the width of the white dwarf eclipse could be affected by the same phenomena that affect the ingress and egress duration. However, in GY~Cnc, the ingress and egress are so short, that any possible effect is smaller than the formal uncertainties on $\Delta \phi_{1/2}$.

In \cite{mcallister2019}, eclipse light curve modelling was used to find $q=0.448^{+0.014}_{-0.021}$ and $\Delta \phi_{1/2}= 0.064 \pm 0.002$. Since $\Delta \phi_{1/2}$ is a function only of $q$ and the inclination $i$, this yields $i=77^{\rm o}.1^{+0.3^{\rm o}}_{-0.2^{\rm o}}$. The orbital period is $P=0.175442399(6)$ day, with the number in brackets being the uncertainty in the last digit. Kepler's third law allows us to write
\begin{equation}
    M_1 = \frac{K_2^3\,P (1+q)^2}{2\pi G \sin^3 i}~.
\end{equation}
Adopting $K_2 = K_{\rm abs} = 280 \pm 2$\,\kms, we find $M_1 = 0.90 \pm 0.03$\,\msun. Unsurprisingly, this is in good agreement with $M_1 = 0.88 \pm 0.02$\,\msun from \cite{mcallister2019} and the less precise measure by \cite{thorstensen2000}  of $M_1 = 0.99 \pm 0.12$\,\msun. It is however, more than 5$\sigma$ different from the \cite{godon2022} white dwarf mass of $0.57 \pm 0.05$\,\msun, derived through careful modelling of the ultraviolet spectrum of GY~Cnc. We therefore conclude that the higher white dwarf mass is correct and recommend the adoption of the system parameters in \cite{mcallister2019} for GY~Cnc.

\section{Conclusions}
Stellar masses in CVs derived from modelling the white dwarf eclipses rely on several assumptions. There is little evidence for most of these assumptions and it is important to keep that in mind when using masses derived in this manner. However, whenever masses derived from light curve modelling have been checked independently, they have proved to be consistent \citep{tulloch09, copperwheat2012, savoury12}. GY~Cnc was the only system for which \cite{godon2022} found a discrepancy between the white dwarf masses obtained from the eclipse light curve and ultraviolet spectroscopy. We show here that the white dwarf mass derived from the eclipse light curve  is the correct one.

Therefore, we conclude that masses in CVs obtained via eclipse light curve modelling are generally reliable. It is beyond the scope of this paper to establish why the analysis of the ultraviolet spectrum by \cite{godon2022} found an anomalously low white dwarf mass. However, we can speculate that it is related to the difficulty of analysing the white dwarf spectrum in the presence of contamination from the accretion disc, boundary layer and iron curtain.

Whilst it is possible that applying eclipse light curve modelling to any single system could produce incorrect masses due to the breakdown of the assumptions given in the introduction, there remains no known example where this has actually occurred. Thus, we recommend caution when drawing conclusions from the parameters of a single system, but believe that conclusions drawn from the wider population or the distribution of component masses are robust.

\section*{Acknowledgements}
This paper is dedicated to the memory of our friend and colleague Tom Marsh who sadly died whilst this paper was in preparation. Without his pioneering techniques in the analysis of binary stars and selfless effort providing analysis tools to the community, this paper---and many others---would never have existed. 
\noindent
PR-G acknowledges support from the Consejería de Economía, Conocimiento y Empleo del Gobierno de Canarias and the European Regional Development Fund (ERDF) under grant with reference ProID2021010132 and ProID2020010104. SPL, SGP and VSD acknowledge the support of the Science and Technology Facilities Council (grant ST/V000853/1). This paper includes observations made in the Observatorios de Canarias del IAC with the WHT operated on the island of La Palma by the IAC in the Observatorio del Roque de los Muchachos and makes use of data obtained from the Isaac Newton Group Archive which is maintained as part of the CASU Astronomical Data Centre at the Institute of Astronomy, Cambridge. For the purpose of open access, the authors has applied a creative commons attribution (CC BY) licence to any author accepted manuscript version arising. 

\section*{Data Availability}
The data underlying this article will be shared on reasonable request to the corresponding author.



\bibliographystyle{mnras}
\bibliography{new_refs, refs, refs2, refs3, refs4}





\bsp	
\label{lastpage}
\end{document}